\documentstyle[aps,multicol,floats,epsfig,amsfonts,pra]{revtex}
\begin{document}
\draft
\title{Quantum statistical properties of the radiation field in a 
cavity with a movable mirror}
\author{C. Brif$^{1,2}$ and A. Mann$^{2}$}
\address{$^{1}$LIGO Project, California Institute of Technology, 
Pasadena, CA 91125 \\
$^{2}$Department of Physics, Technion---Israel Institute
of Technology, Haifa 32000, Israel}
\maketitle

\begin{abstract}
A quantum system composed of a cavity radiation field interacting 
with a movable mirror is considered and quantum statistical 
properties of the field are studied. Such a system can serve in
principle as an idealized meter for detection of a weak classical 
force coupled to the mirror which is modelled by a quantum 
harmonic oscillator. It is shown that the standard quantum 
limit on the measurement of the mirror position arises naturally 
from the properties of the system during its dynamical evolution.
However, the force detection sensitivity of the system falls short
of the corresponding standard quantum limit. We also study the 
effect of the nonlinear interaction between the moving mirror and 
the radiation pressure on the quadrature fluctuations of the 
initially coherent cavity field. 
\end{abstract}


\begin{multicols}{2}

\section{Introduction}

Due to recent experimental efforts to detect gravitational waves 
\cite{GWDexp}, the problem of fundamental limitations on the 
detection sensitivity imposed by the quantum nature of the 
measurement device has become of practical importance. 
The research during the last decade focused mostly
on interferometric schemes in which the relative shift in the
positions of two end mirrors, caused by a gravitational wave,
results in a phase shift between the light beams in the two
arms. This phase shift can be detected by measuring 
the output light. The first analysis of quantum limitations on 
the measurement sensitivity in such an interferometric scheme was
made by Caves \cite{Caves} in the early eighties. According to
this analysis, there are two sources of quantum noise: the 
photon-counting noise and the radiation-pressure noise. When
these two sources of noise are balanced by adjusting the light
intensity, the sensitivity achieves the so-called standard quantum 
limit (SQL). 

More recently, Pace \emph{et al.} \cite{PCW93} studied quantum 
limitations on the interferometer sensitivity using a more
detailed analysis based on a Hamiltonian description of the
interaction of light with the movable end mirrors. Some other works 
\cite{VyMa,MaTo94,FPBHGR94,HeRe94} were devoted to the study of 
quantum noise reduction in simpler optical systems: a light beam 
reflected by a movable mirror or a single Fabry-P\'{e}rot cavity 
with one movable mirror. Assuming high light intensities, the above 
works mostly employed semi-classical calculational techniques, with 
the quantum fluctuations of the light field considered as a small
addition to the large classical amplitude. The quantum theory of
the radiation-pressure fluctuations on a mirror was also studied
in \cite{SLB95}. 

In the present paper we consider a simple but nevertheless 
instructive system, consisting of a cavity radiation field 
interacting with a movable mirror \cite{Moo70,Jan98,Law95}. 
As was shown recently 
\cite{MMT97,BJK97}, such a system can be utilized to generate
a variety of non-classical states (especially, the so-called
Schr\"{o}dinger-cat states) of both the cavity field and the 
mirror. We employ the fully quantum description of the system 
and study quantum statistical properties of the cavity field.
While the photon statistics of the field does not change during 
the interaction, the phase distribution and quadrature 
fluctuations exhibit interesting dynamics.

If the mirror (modelled by a quantum harmonic oscillator) is
driven by an external classical force, this will result in
a phase shift of the cavity mode. We study quantum limitations 
imposed on the detection of the force by the intrinsic quantum 
uncertainty in a measurement of this phase shift. 
It is satisfactory to find that the SQL for the measurement of 
the mirror position naturally emerges from the dynamical 
properties of the system. On the other hand, the sensitivity
of the force detection in this scheme does not reach the
corresponding SQL. We also show that analytical approximations
made to describe phase properties of the field should work
very well for light intensities needed to achieve
the SQL in appropriate experimental setups.

\section{The standard quantum limit in the simplest optomechanical
sensor}

As is well known, the SQL arises due to the Heisenberg uncertainty 
relation for the position and momentum of the movable mirror.
This limit can be understood conceptually by considering
a simple measurement scheme consisting of a movable mirror
(modelled, for example, by a linear harmonic oscillator of mass
$m$ and angular frequency $\omega_m$) and a monochromatic light 
beam of angular frequency $\omega_0$. A shift $z$ in the mirror
position, caused by an external classical force (e.g., by a 
gravitational wave), results in a phase shift 
$\phi = 2 \omega_0 z/c$ of the reflected light beam (here $c$ 
is the velocity of light). Such a scheme is referred to as an 
optomechanical sensor. The experimental progress 
towards the observation of quantum limits in such a sensor was 
reported recently in \cite{Tit99}. For a recent theoretical
treatment of such a scheme, including a detailed analysis of
various sources of noise, see \cite{JTWS99}.

If the incident light beam is in a coherent state with mean 
photon number $\bar{N}$, the phase uncertainty is 
$\Delta \phi \simeq 1/(2 \bar{N}^{1/2})$ (provided that
$\bar{N} \gg 1$). This phase-fluctuation noise results in the 
uncertainty 
$(\Delta z)_{\mathrm{pf}} \simeq c/(4 \omega_0 \bar{N}^{1/2})$ 
in the measurement of the mirror position.
Increasing the light intensity, one can suppress the 
phase-fluctuation noise, but this leads to increased
fluctuations in the radiation pressure which on the average 
transfers the momentum $(2 \hbar \omega_0/c) \bar{N}$ to the 
mirror. These radiation-pressure fluctuations result in the 
mirror position uncertainty 
$(\Delta z)_{\mathrm{rp}} \simeq 2 \hbar \omega_0 \bar{N}^{1/2}
/(m c \omega_m)$. 
Optimizing the sum
$(\Delta z)^2 = (\Delta z)_{\mathrm{pf}}^2 + 
(\Delta z)_{\mathrm{rp}}^2$ 
as a function of $\bar{N}$, one finds the uncertainty 
$(\Delta z)_{\mathrm{opt}}$ which is $\sqrt{2}$ times the SQL
for a harmonic oscillator,
\begin{equation}
  \label{eq:zsql}
(\Delta z)_{\mathrm{SQL}} = \sqrt{\hbar/(2 m \omega_m)} .
\end{equation}
The optimum mean number of photons is
\begin{equation}
  \label{eq:nopt}
\bar{N}_{\mathrm{opt}} = m \omega_m c^2/(8 \hbar \omega_0^2) .
\end{equation}

In many applications (e.g., for detection of gravitational waves) 
it makes sense to consider what limitations apply due to the quantum 
nature of the measuring apparatus on the detection of a weak classical 
force coupled to the oscillator \cite{CTDSZ80}. Consider a constant 
force $F$ that acts on the mirror during time $t$ which is much
shorter than the oscillator period $2\pi/\omega_m$. In order to
detect this force, the momentum $p_F \simeq F t$ it transfers to 
the mirror should be larger than the SQL for the measurement of
the oscillator momentum, 
$(\Delta p)_{\mathrm{SQL}} = (\hbar m \omega_m /2)^{1/2}$.
Correspondingly, the minimum detectable force is
\begin{equation}
  \label{eq:Fsql}
F_{\mathrm{SQL}} = t^{-1} (\Delta p)_{\mathrm{SQL}}
= t^{-1} \sqrt{\hbar m \omega_m /2} .
\end{equation}

It was argued \cite{Unr83,BoSh84,JaRe90,BrKh92} that the SQL 
on the detection of a weak classical force (which causes the 
displacement of the movable mirror in the sensor) can be 
surpassed using correlations between the phase-fluctuation noise 
and the radiation-pressure noise. In particular, it was 
predicted \cite{VyMa,MaTo94,FPBHGR94,HeRe94} that amplitude-phase 
correlations, created by the nonlinear interaction between the 
movable mirror and the radiation pressure, will produce squeezing 
in the reflected beam. Consequently, it was proposed \cite{VyMa}
to use homodyne detection of the squeezed quadrature in order
to compensate for the radiation-pressure noise, thereby making 
it possible to overcome the SQL with high enough light intensity.

\section{The model}

The solution of the wave equation for the radiation field in a
cavity with moving boundaries is in general a very complicated 
problem even in the classical case \cite{Moo70,Jan98}. However,
in the adiabatic approximation, when the amplitude and frequency
of the mirror motion are much smaller than the wavelength and 
frequency of the cavity mode, the dynamics of the system can
be described by an effective Hamiltonian \cite{Law95}. Of course,
all phenomena associated with the dynamical Casimir effect
(e.g., the resonant generation of photons) are neglected in this
description.

The free Hamiltonian for the system of the radiation field in
a linear cavity with one movable mirror is
\begin{equation}
  \label{eq:Hfree}
H_0 = H_c + H_m = 
\hbar \omega_c a^{\dagger} a + \hbar \omega_m b^{\dagger} b , 
\end{equation}
where $\omega_c$ and $a$ are the angular frequency and the boson 
annihilation operator of the cavity mode, while $\omega_m$ and $b$ 
are the angular frequency and the annihilation operator of the 
movable mirror modelled by a quantum harmonic oscillator. 
For a cavity of free length $L$, one has 
$\omega_c = \pi (c/L) n$, $n \in \mathbb{N}$.
If the mirror displacement from the equilibrium position is $z$,
then the cavity length changes $L \rightarrow L+z$, and the
cavity angular frequency changes as
\begin{equation}
  \label{eq:omc-change}
\omega_c \rightarrow \pi \frac{c}{L+z} n \approx 
\omega_c \left( 1 - \frac{z}{L} \right) .
\end{equation}
Therefore, the effective interaction Hamiltonian reads
\begin{equation}
  \label{eq:Hint}
H_{\mathrm{int}} = - \hbar \omega_c \frac{z}{L} a^{\dagger} a
= - \hbar g a^{\dagger} a (b + b^{\dagger}) ,
\end{equation}
where
\begin{eqnarray}
& & z = \left( \frac{\hbar}{2 m \omega_m} \right)^{1/2} 
(b + b^{\dagger}) , \\ 
& & g = \frac{\omega_c}{L} 
\left( \frac{\hbar}{2 m \omega_m} \right)^{1/2} .
\end{eqnarray}
Here, $m$ is the mass of the movable mirror.
If, in addition, we assume that the mirror is driven by a 
classical external time-dependent force $F(t)$, the corresponding 
part of the Hamiltonian is given by
\begin{eqnarray}
  \label{eq:HF}
& & H_F = - F(t) z = - \hbar f(t) (b + b^{\dagger}) , \\
& & f(t) = (2 m \omega_m \hbar)^{-1/2} F(t) .
\end{eqnarray}
Then the total Hamiltonian 
$H_{\mathrm{tot}} = H_0 + H_{\mathrm{int}} + H_F$ reads
\begin{equation}
  \label{eq:Htot}
H_{\mathrm{tot}} = 
\hbar \omega_c a^{\dagger} a + \hbar \omega_m b^{\dagger} b 
- \hbar g a^{\dagger} a (b + b^{\dagger})
- \hbar f(t) (b + b^{\dagger}) .
\end{equation}

It is known that the time evolution operator for this system can be 
written in a closed form \cite{MMT97,BJK97}. One can easily verify 
that this evolution operator is given by
\begin{equation}
  \label{eq:Ut1}
U(\tau) = \exp( - i r a^{\dagger} a \tau)
\exp[ - i b^{\dagger} b \tau + i (k a^{\dagger} a + \lambda)
(b + b^{\dagger}) \tau ] .
\end{equation}
With some algebra \cite{BJK97}, this operator can be rewritten
as
\begin{eqnarray}
  \label{eq:Ut2}
U(\tau) & = & \exp( - i r a^{\dagger} a \tau)
\exp[ i \mu (k a^{\dagger} a + \lambda)^2 ] \nonumber \\
& & \times \exp[ (k a^{\dagger} a + \lambda) 
(\eta b^{\dagger} - \eta^{\ast} b) ]
\exp( - i b^{\dagger} b \tau ) .
\end{eqnarray}
Here, $\tau = \omega_m t$ is the scaled time and other scaled
parameters are $r = \omega_c / \omega_m$ and $k = g / \omega_m$.
Also, $\lambda$, $\mu$, and $\eta$ are dimensionless functions 
of time:
\begin{eqnarray}
& & \lambda(\tau) = \frac{1}{\tau} \int_{0}^{\tau/\omega_m}
f(t') d t' , \\
& & \mu(\tau) = \tau - \sin \tau , \\
& & \eta(\tau) = 1 - \exp( - i \tau) .
\end{eqnarray}
An interesting feature of this system is that for times
$\tau = 2 \pi, 4 \pi, \ldots$ (i.e., after a full cycle of the
mirror oscillatory motion), one finds $\eta = 0$, and the field 
and mirror subsystems become fully disentangled (if they were
uncorrelated initially). Furthermore, at these moments, the 
mirror returns to its initial state. This effect is quite general
as it does not depend on the initial states of the subsystems.
Another interesting property is that the photon statistics of the
cavity field does not change during the interaction.

The existence of the closed form (\ref{eq:Ut1}) for the
evolution operator can be understood if we note that the
Hamiltonian (\ref{eq:Htot}) is given by a linear combination of 
operators which close an algebra. First, recall that a solvable
Lie algebra (the so-called oscillator algebra) is closed by the
four operators $\{N_b, b, b^{\dagger}, I \}$, where 
$N_b = b^{\dagger} b$ is the number operator and $I$ is the
identity operator. Non-vanishing commutation relations are
\begin{equation}
  \label{eq:osc-alg1}
[N_b , b] = - b , \hspace{7mm} 
[N_b , b^{\dagger}] = b^{\dagger} , \hspace{7mm}
[b , b^{\dagger}] = I .
\end{equation}
Alternatively, one can choose for the oscillator algebra the fully 
Hermitian basis $\{N_b, X_b, Y_b, I \}$, where 
$X_b = 2^{-1/2} (b + b^{\dagger})$ and 
$Y_b = 2^{-1/2} (b - b^{\dagger})/ i $ are the scaled position
and momentum operators. Non-vanishing commutation relations are
\begin{equation}
  \label{eq:osc-alg2}
[N_b , X_b] = - i Y_b , \hspace{7mm} 
[N_b , Y_b] = i X_b , \hspace{7mm}
[X_b , Y_b] = i I .
\end{equation}
It is straightforward to see that the oscillator algebra can also
be constructed using two bosonic modes $a$ and $b$. The
Hermitian basis is given in the two-mode realization by
$\{N_b, N_a X_b, N_a Y_b, N_a^2 \}$, where $N_a = a^{\dagger} a$
is the number operator for the mode $a$, and non-vanishing 
commutation relations are
\begin{eqnarray}
& & [N_b , N_a X_b] = - i N_a Y_b , \hspace{7mm} 
[N_b , N_a Y_b] = i N_a X_b , \nonumber \\
& & [N_a X_b , N_a Y_b] = i N_a^2 .
  \label{eq:osc-alg3}
\end{eqnarray}
More generally, it is also possible to consider a solvable Lie 
algebra which is closed by the eight operators 
$\{N_b, N_a X_b, N_a Y_b, X_b, Y_b, N_a^2, N_a, I \}$ with
non-vanishing commutation relations given by the combination of
(\ref{eq:osc-alg2}), (\ref{eq:osc-alg3}) and
\begin{equation}
[N_a X_b , Y_b] = [X_b , N_a Y_b] = i N_a .
\end{equation}
Since the Hamiltonian (\ref{eq:Htot}) is given by a linear 
combination of the operators $N_a$, $N_b$, $N_a X_b$, and $X_b$,
the Baker-Campbell-Hausdorff theorem \cite{Mag54} assures the 
existence of a closed expression for the time evolution operator.

\section{The evolution of coherent states}

Let us consider a situation in which both the cavity field and the
mirror oscillator are initially prepared in coherent states
with coherent amplitudes $\alpha$ and $\beta$, respectively:
\begin{equation}
  \label{eq:Psi0}
| \Psi(0) \rangle = |\alpha\rangle_c \otimes |\beta\rangle_m
= e^{-|\alpha|^2 /2} \sum_{n=0}^{\infty} 
\frac{\alpha^n}{\sqrt{n!}} |n\rangle_c \otimes |\beta\rangle_m .
\end{equation}
In fact, a quite reasonable choice is to assume that the mirror 
is initially in the vacuum state (i.e., $\beta = 0$), but first 
we will consider an arbitrary $\beta$ for the sake of generality. 
Acting on the initial state (\ref{eq:Psi0}) with the evolution 
operator (\ref{eq:Ut2}), one obtains:
\begin{eqnarray}
| \Psi(\tau) \rangle & = & e^{-|\alpha|^2 /2} \sum_{n=0}^{\infty} 
\frac{(\alpha e^{- i r \tau})^n}{\sqrt{n!}} 
e^{ i \mu (k n + \lambda)^2 } |n\rangle_c \nonumber \\
& & \otimes 
\left[ e^{ (k n + \lambda) (\eta b^{\dagger} - \eta^{\ast} b) }
|\beta e^{- i \tau} \rangle_m \right] .
\end{eqnarray}
Utilizing properties of the displacement operators \cite{Gla63},
the state of the system may be written as (we omit an unimportant 
overall phase factor)
\begin{eqnarray}
| \Psi(\tau) \rangle & = & e^{-|\alpha|^2 /2} \sum_{n=0}^{\infty} 
\frac{[\alpha e^{- i \zeta(\tau)}]^n }{\sqrt{n!}} \nonumber \\
& & \times e^{ i \mu(\tau) k^2 n^2 } |n\rangle_c \otimes 
|\gamma_n (\tau)\rangle_m .
  \label{eq:Psi-tau}
\end{eqnarray}
In the reference frame rotating with the field frequency $\omega_c$,
the phase $\zeta(\tau)$ is given by
\begin{equation}
  \label{eq:zeta}
\zeta(\tau) = 2 k \lambda(\tau) \mu(\tau) + 
k \mathrm{Im}\, [\beta \eta(\tau)] .
\end{equation}
In Eq.~(\ref{eq:Psi-tau}), $|\gamma_n (\tau)\rangle$ is a 
coherent state with the amplitude 
\begin{equation}
  \label{eq:gamma}
\gamma_n (\tau) = \beta e^{- i \tau} + 
[k n + \lambda(\tau)] \eta(\tau) .
\end{equation}
For $\tau = 2\pi, 4\pi, \ldots$, the dependence of this
amplitude on $n$ disappears, and the system returns to an
uncorrelated product state.

In what follows we will be interested in quantum statistical
properties of the cavity field. These properties can be calculated
from the reduced density matrix, obtained by tracing out the
mirror degrees of freedom,
\begin{eqnarray}
  \label{eq:rho-c}
\rho_c (\tau) & = & {\mathrm{Tr}}_m \left\{ |\Psi(\tau) \rangle
\langle \Psi(\tau)| \right\} \nonumber \\
& = & e^{-|\alpha|^2} \sum_{n=0}^{\infty} \sum_{n'=0}^{\infty}
\frac{ (\alpha e^{ i \zeta})^n 
(\alpha^{\ast} e^{- i \zeta})^{n'} }{\sqrt{n! n'!}} \nonumber \\
& & \times e^{ i \mu k^2 (n^2 -n'^2) }
\langle \gamma_{n'} | \gamma_n \rangle
| n\rangle_c \, {}_c\langle n' | ,
\end{eqnarray}
where
\begin{eqnarray}
\langle \gamma_{n'} | \gamma_n \rangle & = & 
\exp\left( \gamma_{n'}^{\ast} \gamma_n - 
\mbox{$\frac{1}{2}$} |\gamma_n|^2 - 
\mbox{$\frac{1}{2}$} |\gamma_{n'}|^2  \right) \\
& = & \exp\left[ - \dot{\mu} k^2 (n-n')^2 
\right]  \hspace{5mm} \mathrm{for} \ \ \beta = 0 .
\label{eq:scalprod}
\end{eqnarray}
Note that $\dot{\mu} = \frac{1}{2} |\eta|^2 = 1 - \cos \tau$.
Obviously, the Poissonian photon statistics of the cavity field 
is preserved during the evolution.

\section{Phase properties of the cavity field}

As can be seen from equations (\ref{eq:Psi-tau}) and (\ref{eq:zeta}),
the external force acting on the mirror produces a phase shift of
the cavity mode. At least in principle, this phase shift can be 
measured, thereby revealing information about the classical force.
Accuracy of such a measurement will inevitably depend on the phase
properties of the radiation mode, because an externally induced 
phase shift is detectable only if it is larger than the intrinsic 
phase uncertainty of the field.

\subsection{Canonical and realistic phase distributions}

In quantum mechanics, one can describe a measurement by the
corresponding positive operator-valued measure (POVM) 
\cite{Hel76}. The so-called canonical POVM for an idealized
phase measurement is given by \cite{ShSh91,BBA94}
\begin{equation}
  \label{eq:POVM-def}
\Pi_{\mathrm{can}}(\theta) = 
\frac{1}{2\pi} |\theta\rangle \langle\theta| ,
\hspace{7mm} |\theta\rangle = \sum_{n=0}^{\infty}
e^{ i n \theta } |n\rangle .
\end{equation}
The corresponding phase distribution for the cavity field is
\begin{equation}
  \label{eq:Pt-def}
P_{\mathrm{can}}(\theta) = 
{\mathrm{Tr}}[ \Pi_{\mathrm{can}}(\theta) \rho_c ]
= (2\pi)^{-1} \langle\theta | \rho_c | \theta\rangle .
\end{equation}
Substituting expression (\ref{eq:rho-c}) for $\rho_c$ into 
(\ref{eq:Pt-def}), we obtain
\begin{eqnarray}
  \label{eq:Pt1}
P_{\mathrm{can}}(\theta) & = & \frac{e^{-|\alpha|^2}}{2\pi}
\sum_{n=0}^{\infty} \sum_{n'=0}^{\infty}
\frac{ |\alpha|^{n+n'} }{\sqrt{n! n'!}}
e^{- i (\theta-\zeta-\varphi_{\alpha}) (n-n')} \nonumber \\
& & \times e^{ i \mu k^2 (n^2 -n'^2) }
\langle \gamma_{n'} | \gamma_n \rangle ,
\end{eqnarray}
where $\varphi_{\alpha} = \arg(\alpha)$. For $\beta = 0$, we can 
use expression (\ref{eq:scalprod}) for 
$\langle \gamma_{n'} | \gamma_n \rangle$. Then the double summation 
in Eq.~(\ref{eq:Pt1}) can be rearranged in such a way that
$P_{\mathrm{can}}(\theta)$ takes the form of a Fourier series:
\begin{equation}
  \label{eq:Pt2}
P_{\mathrm{can}}(\theta) = \frac{1}{2\pi} 
\sum_{q = -\infty}^{\infty} e^{-\dot{\mu} k^2 q^2} 
{\cal A}_q e^{ i (\theta-\zeta-\varphi_{\alpha}) q} ,
\end{equation}
where
\begin{eqnarray}
  \label{eq:Aq}
& & {\cal A}_q = e^{-|\alpha|^2} \xi_q^{|q|}
\sum_{n=0}^{\infty} \frac{\xi_q^{2n}}{\sqrt{n! (n+|q|)!}} , \\
& & \xi_q = |\alpha| \exp(- i \mu k^2 q) .
  \label{eq:xiq}
\end{eqnarray}

The problem of the quantum description of optical phase and the 
problem of its measurement were widely discussed during the last 
decade (see, e.g., reviews \cite{TMG96,PeBa97}). 
However, existence of an experimental procedure for 
phase measurement with the canonical POVM (\ref{eq:POVM-def}) 
is still an open question. On the other hand, a possible way to 
measure the phase of the radiation field mode is by means of 
heterodyne detection (see, e.g., \cite{WiMi93}). 
Relating to our model, a possible experimental procedure might be 
as follows. At some moment $t$ one stops the interaction between 
the field and the movable mirror (by fixing the mirror position) 
and lets the radiation leak from the cavity. The output field is 
detected using the heterodyne scheme, which employs a strong local 
oscillator highly detuned from the signal. The two Fourier 
components of the photocurrent are proportional to the quadratures 
of the cavity field at the moment $t$, which can be used to 
determine the phase. The POVM for this phase measurement is
\begin{equation}
  \label{eq:POVM-Q}
\Pi_Q (\theta) = \frac{1}{\pi} \int_{0}^{\infty} 
| A \rangle \langle A |\, r d r ,
\hspace{7mm} A = r e^{ i \theta} ,
\end{equation}
where $| A \rangle$ is a coherent state with amplitude $A$.
The corresponding phase distribution is obtained by integrating
the $Q$ distribution function,
\begin{equation}
  \label{eq:Qfunction}
Q(A) = \frac{1}{\pi} \langle A | \rho_c | A \rangle 
\end{equation}
over the radial coordinate in the complex plane:
\begin{equation}
  \label{eq:PQt-def}
P_Q(\theta) = {\mathrm{Tr}}[ \Pi_Q (\theta) \rho_c ]
= \int_{0}^{\infty} Q(r e^{ i \theta}) r d r .
\end{equation}
For the cavity field mode with $\rho_c$ of equation
(\ref{eq:rho-c}), we obtain
\begin{eqnarray}
P_Q(\theta) & = & \frac{e^{-|\alpha|^2}}{2\pi}
\sum_{n=0}^{\infty} \sum_{n'=0}^{\infty}
\Gamma\left( \frac{n+n'+2}{2} \right)
\frac{ |\alpha|^{n+n'} }{n! n'!} \nonumber \\
& & \times e^{- i (\theta-\zeta-\varphi_{\alpha}) (n-n')} 
e^{ i \mu k^2 (n^2 -n'^2) }
\langle \gamma_{n'} | \gamma_n \rangle .
  \label{eq:PQt1}
\end{eqnarray}
We see that the distributions $P_{\mathrm{can}}(\theta)$ and 
$P_Q(\theta)$ differ by the factor
\begin{equation}
\frac{1}{\sqrt{n! n'!}} \Gamma\left( \frac{n+n'+2}{2} \right)
\end{equation}
in the double sum over $n$ and $n'$. Correspondingly, $P_Q(\theta)$ 
is usually slightly wider than $P_{\mathrm{can}}(\theta)$.
(It has been argued \cite{LVBP95} that this widening is related 
to contamination by external noise during the heterodyne detection.)
For $\beta = 0$, the rearrangement of the double summation in 
(\ref{eq:PQt1}) gives
\begin{equation}
  \label{eq:PQt2}
P_Q(\theta) = \frac{1}{2\pi} \sum_{q = -\infty}^{\infty}
e^{-\dot{\mu} k^2 q^2} {\cal B}_q 
e^{ i (\theta-\zeta-\varphi_{\alpha}) q} ,
\end{equation}
where
\begin{equation}
  \label{eq:Bq}
{\cal B}_q = e^{-|\alpha|^2} \xi_q^{|q|}
\sum_{n=0}^{\infty} 
\frac{\Gamma(n+ \frac{1}{2}|q| + 1)}{\Gamma(n+ |q| + 1)}
\frac{\xi_q^{2n}}{n!} ,
\end{equation}
and $\xi_q$ is given by Eq.~(\ref{eq:xiq}).
Note that ${\cal B}_0 = 1$. The summation in (\ref{eq:Bq}) is 
proportional to the Kummer series \cite{Erd53}, so we can 
write
\begin{equation}
  \label{eq:Bq-CHGF}
{\cal B}_q = e^{-|\alpha|^2} \xi_q^{|q|} 
\frac{\Gamma(\frac{1}{2}|q| + 1)}{\Gamma(|q| + 1)}
\Phi(\mbox{$\frac{1}{2}$} |q| + 1, |q|+1; \xi_q^2) .
\end{equation}
Here, $\Phi(a,b;x)$ is the Humbert symbol for the confluent 
hypergeometric function (the Kummer function). Using properties 
of special functions, ${\cal B}_q$ can also be written as
\begin{eqnarray}
{\cal B}_q & = & \frac{\sqrt{\pi}}{2} \exp( -|\alpha|^2 )
\exp(\xi_q^2 /2)\, \xi_q \nonumber \\
& & \times \left[ I_{(|q|-1)/2} (\xi_q^2 /2)
+ I_{(|q|+1)/2} (\xi_q^2 /2) \right] ,
  \label{eq:Bq-BF}
\end{eqnarray}
where $I_{\nu}(x)$ is the modified Bessel function of the
first kind.
In the strong-field limit $|\alpha|^2 \gg 1$, we can use the
asymptotic expansion for the confluent hypergeometric function
or for the modified Bessel function to obtain (for fixed $q$):
\begin{equation}
  \label{eq:Bq-asy}
{\cal B}_q \approx \exp(-|\alpha|^2 + \xi_q^2)
\left( 1 - \frac{q^2}{4 \xi_q^2} + \cdots \right) .
\end{equation}
For high field intensities, the behavior of both phase 
distributions is qualitatively very similar. In what follows
we will consider the $P_Q(\theta)$ distribution, because of
two reasons: (i) phase measurements with the corresponding POVM 
$\Pi_Q(\theta)$ are experimentally feasible and (ii) the existence 
of a closed expression for ${\cal B}_q$ gives a calculational 
advantage. 

Let us assume the phase range $-\pi \leq \theta < \pi$. Then
moments of the phase distribution are determined by
\begin{equation}
  \label{eq:moments}
\bar{\theta} = \int_{-\pi}^{\pi} \theta P_Q(\theta) d \theta ,
\hspace{7mm} \overline{\theta^2} = \int_{-\pi}^{\pi} \theta^2 
P_Q(\theta) d \theta .
\end{equation}
A simple calculation gives
\begin{equation}
  \label{eq:mean}
\bar{\theta} = 2 \sum_{q=1}^{\infty} \frac{1}{q} 
e^{-\dot{\mu} k^2 q^2} {\mathrm{Im}} \left[ {\cal B}_q 
e^{- i (\zeta+\varphi_{\alpha}+\pi) q} \right] ,
\end{equation}
\begin{equation}
  \label{eq:sqmean}
\overline{\theta^2} = \frac{\pi^2}{3} + 4 \sum_{q=1}^{\infty} 
\frac{1}{q^2} e^{-\dot{\mu} k^2 q^2}  {\mathrm{Re}} \left[ 
{\cal B}_q e^{- i (\zeta+\varphi_{\alpha}+\pi) q} \right] .
\end{equation}
Using these expressions, we can calculate the phase
uncertainty, 
$\Delta \theta = ( \overline{\theta^2} - \bar{\theta}^2 )^{1/2}$.
This can be done numerically, but some important information may
also be deduced from an analytical approximation.

\subsection{Analytical approximations and numerical results}

In our considerations we neglected losses which occur in a realistic
case due to the relaxation of the cavity field and the dissipation
of the mirror oscillator. For a cavity with a very high finesse and
a mirror oscillator with a very high quality factor, neglecting 
the losses can be a fair approximation for short interaction times,
say, for $t < 10^{-3}$ s. Therefore, for low-frequency oscillators 
(with $\omega_m \sim 2\pi$ Hz, as in the LIGO interferometric
gravitational-wave detectors \cite{GWDexp}), it is reasonable to 
consider $\tau < 10^{-2}$. For such small values of $\tau$ we can 
use approximate expressions for $\mu(\tau)$ and $\dot{\mu}(\tau)$.
The leading term in $\dot{\mu}(\tau)$ is $\tau^2/2$, so the
terms in the series (\ref{eq:PQt2}) are weighted by the prefactor
$\exp(-k^2 \tau^2 q^2/2)$. Correspondingly, the contribution of
terms with $q$ much larger than $q_m = (k \tau)^{-1}$ is rather
insignificant. Then, for high field intensities (i.e., $|\alpha|$ 
much larger than $q_m$ and $|\alpha|^2 \gg 1$),
we may approximate ${\cal B}_q$ by just the leading term in 
its asymptotic expansion (\ref{eq:Bq-asy}). With all these
assumptions, we derive
\begin{equation}
  \label{eq:PQt-appr}
P_Q(\theta) \approx \frac{1}{2\pi} \sum_{q = -\infty}^{\infty}
\exp[ - \mbox{$\frac{1}{2}$} \sigma^2 q^2 ]
\exp[ i (\theta - \tilde{\theta}) q ] ,
\end{equation}
\begin{equation}
\sigma = k \tau \sqrt{1 + \varepsilon_1} , \hspace{7mm}
\tilde{\theta} = \zeta + \varphi_{\alpha} + \varepsilon_2 ,
\end{equation}
\begin{equation}
\varepsilon_1 = \frac{k^2 \tau^4 |\alpha|^2}{9} - 
\frac{\tau^2}{12} + \frac{\tau^4}{360} , \hspace{7mm}
\varepsilon_2 = \frac{k^2 \tau^3 |\alpha|^2}{3} .
\end{equation}
We see that the phase distribution $P_Q(\theta)$ is approximated 
by a Fourier series with Gaussian coefficients, 
\begin{equation}
  \label{eq:Fcoeff}
\exp[ - \mbox{$\frac{1}{2}$} \sigma^2 q^2 ]
= \int_{-\pi}^{\pi} P_Q(\theta) 
e^{- i (\theta - \tilde{\theta}) q} d \theta .
\end{equation}
Though we did not find a closed expression for the series
(\ref{eq:PQt-appr}), a further approximation may be made. 
It is clear that the width of the distribution is of the order of 
$\sigma$ and it is centered near $\tilde{\theta}$. 
When the distribution is very narrow (i.e., $\sigma \ll 1$) and its 
center is far from the boundaries of the interval $[-\pi, \pi)$ 
(in comparison with the distribution width $\sigma$), it 
is possible to extend the integration range in (\ref{eq:Fcoeff}) 
from $-\infty$ to $+\infty$ without changing the result 
significantly. This is equivalent to the replacement of the
Fourier series in (\ref{eq:PQt-appr}) by the Fourier integral.
Formally, we can apply the Poisson summation formula (see, e.g.,
\cite{Stak67}) to the series (\ref{eq:PQt-appr}) to
obtain 
\begin{equation}
  \label{eq:PQt-PSF}
P_Q(\theta) \approx (2 \pi \sigma^2)^{-1/2} \sum_{m=-\infty}^{\infty} 
\exp\left[- \mbox{$\frac{1}{2}$} (\theta - \tilde{\theta} - 2 m \pi)^2
/ \sigma^2 \right] .
\end{equation}
For $\sigma \ll 1$, the series (\ref{eq:PQt-PSF}) converges much
faster than (\ref{eq:PQt-appr}). If only the leading term in
(\ref{eq:PQt-PSF}) is left, the phase distribution is approximated
by just the Gaussian,
\begin{equation}
  \label{eq:PQt-Gauss}
P_Q(\theta) \approx (2 \pi \sigma^2)^{-1/2} 
\exp\left[- \mbox{$\frac{1}{2}$} (\theta - \tilde{\theta})^2/
\sigma^2 \right] ,
\end{equation}
centered about $\tilde{\theta}$ and with the dispersion $\sigma$. 

The accuracy of our approximations may be verified by 
calculating numerically the mean phase $\bar{\theta}$ and the 
phase uncertainty $\Delta \theta$ from the distributions 
(\ref{eq:PQt2}) and (\ref{eq:PQt-appr}). For simplicity, we took 
$\zeta = \varphi_{\alpha} = 0$ in these calculations.
Figure~1 shows how the phase uncertainty 
$\Delta \theta$ depends on the intensity of the cavity field.
The approximation (\ref{eq:PQt-appr}) works extremely well
for sufficiently large field intensities. The Gaussian 
approximation (\ref{eq:PQt-Gauss}) is also excellent. In
principle, $\Delta \theta$ obtained from (\ref{eq:PQt-appr})
coincides with $\sigma$ except for the neighborhood of points 
where the mean phase is $\pm \pi\ \mathrm{mod}\ 2 \pi$. The 
corresponding deterioration of the phase sensitivity is an 
artifact of the definition of the phase range from 
$-\pi$ to $\pi$. The dependence of the mean phase $\bar{\theta}$
on $|\alpha|$ is shown in Fig.~2. It is seen that
the value $\tilde{\theta}$ obtained in the Gaussian approximation
differs very insignificantly from the exact value of 
$\bar{\theta}$. One can also verify that the peaks in 
Fig.~1 appear exactly at the points for which
$\bar{\theta}$ in Fig.~2 approaches $\pi$.
Figure~3 shows the dependence of the phase uncertainty 
$\Delta \theta$ on the scaled measurement time $\tau$. Once again,
the approximations work very well.

We should note that the value of $k$ we used ($k = 7.0$) has 
no special physical meaning. In fact, this value is much larger 
than one would expect in an experiment. We chose such a large $k$ 
only in order to deal with smaller values of $|\alpha|$ and 
thereby avoid overflows in numerical calculations. 
However, this artificial scaling does not distort the physical 
behavior of the system. Let us consider what will happen for a 
realistic value of $k$ (e.g., $k \sim 6\times 10^{-8}$ for the 
LIGO detectors). Then one obtains a much larger value of 
$q_m = (k \tau)^{-1}$ (e.g., $q_m \sim 3\times 10^{9}$ for the 
LIGO detectors with the cavity storage time about $10^{-3}$ s). 
Therefore, our approximations might be used, but for very large 
values of $|\alpha|$. In fact, a simple calculation gives
$q_m^2 = (4 L/c t)^2 \bar{N}_{\mathrm{opt}}$, where $t$ is
the measurement time and $\bar{N}_{\mathrm{opt}}$ is the mean
photon number needed to achieve the SQL in an optomechanical 
sensor, as given by Eq.~(\ref{eq:nopt}). With the LIGO
parameters, one obtains 
$\bar{N}_{\mathrm{opt}} \sim 2.5\times 10^{21}$,
so our approximations will work perfectly for the mean
photon numbers $\bar{N} = |\alpha|^2$ of the order of
$\bar{N}_{\mathrm{opt}}$ and even smaller.

\section{Quantum limitations}

The principal result is that, in the region of small $\tau$, the
phase uncertainty $\Delta \theta$ is limited from below by the 
value $k \tau$ (in fact, this limit is slightly higher due to the 
factor $\sqrt{1 + \varepsilon_1}$ in $\sigma$, but this small 
correction is not important for understanding the basic behavior 
of the system). Now, let us consider what this phase uncertainty
means regarding the measurement of the mirror shift. A phase shift
$\phi$ produced due to the mirror position shift $z$ is given
by $\phi \simeq (2 B \omega_c /c) z$, where 
\begin{equation}
B = \frac{t c}{2 L} = \frac{\tau c}{2 \omega_m L}
\end{equation}
is the effective number of bounces by light on the movable mirror
during the measurement time $t$. If the uncertainty of the phase
shift measurement is $\Delta \theta$, the corresponding 
uncertainty of the mirror position shift is
\begin{equation}
\Delta z \simeq \frac{c}{2 B \omega_c} \Delta \theta
= \frac{\omega_m L}{\omega_c \tau} \Delta \theta .
\end{equation}
Since in our model the phase uncertainty $\Delta \theta$ is limited 
from below by $k \tau$, the corresponding limit on the position shift 
uncertainty is
\begin{equation}
(\Delta z)_{\mathrm{min}} \simeq 
\frac{\omega_m L}{\omega_c \tau} k \tau 
= \frac{g L}{\omega_c} = \sqrt{\frac{\hbar}{2 m \omega_m}} ,
\end{equation}
which is exactly the SQL for the position measurement of a harmonic 
oscillator.

The limitation on the detection of an external classical force 
$F(t)$ will depend, of course, not only on the uncertainty 
$\Delta \theta$ of the phase measurement, but also on the value of 
the phase shift $\zeta_F = 2 k \lambda \mu$ produced by this force. 
The corresponding signal-to-noise ratio is
\begin{equation}
\frac{S}{N} = \frac{\zeta_F (\tau)}{\Delta \theta}
= \frac{2 k \lambda(\tau) \mu(\tau)}{\Delta \theta} .
\end{equation}
For $\tau \ll 1$, taking for $\Delta \theta$ its lower limit 
$k \tau$, we obtain
\begin{equation}
\frac{S}{N} \simeq \frac{k \lambda(\tau) \tau^3 /3}{k \tau} =
(2 \hbar m \omega_m)^{-1/2}\, \frac{\tau}{3} 
\int_{0}^{\tau/\omega_m} F(t') d t' .
\end{equation}
Clearly, the sensitivity of the force detection depends on the 
total phase shift produced by the force during the time 
$t = \tau/\omega_m$. This result just reflects the fact that in 
our model the information about the system is inferred from a 
``single-shot'' phase measurement at the time $t$, and not from 
a continuous monitoring of the system. In the case of a force 
$F$ which is constant during the time $t$, we obtain
\begin{equation}
\frac{S}{N} \simeq (2 \hbar m \omega_m)^{-1/2} 
\frac{F \tau^2}{3 \omega_m} .
\end{equation}
The minimum detectable force $F_{\mathrm{min}}$ is determined
from the condition $S/N = 1$. Then we obtain
\begin{equation}
\label{eq:Fmin}
F_{\mathrm{min}} \simeq \sqrt{\frac{18 \hbar m}{\omega_m t^4}} 
= \frac{6}{\tau} F_{\mathrm{SQL}} .
\end{equation}
So, with the method presented here, the minimum detectable force
is quite far from the SQL. This result shows that the genuine
sensitivity in a specific scheme can differ significantly from
the commonly accepted SQL for an abstract oscillator. 

\section{Quantum fluctuations of the field quadratures}

It is well known (see, e.g., \cite{Wis95}) that homodyne detection
with a strong local oscillator can be used to measure the rotated
field quadrature
\begin{equation}
X_{\varphi} = a e^{- i \varphi} + 
a^{\dagger} e^{ i \varphi} .
\end{equation}
The angle $\varphi$ is the phase of the local oscillator.
A quantum state of the signal field $a$ exhibits the phenomenon
of squeezing if the uncertainty 
$\Delta X_{\varphi} = (\langle X_{\varphi}^2 \rangle
- \langle X_{\varphi} \rangle^2)^{1/2}$
is below its vacuum value 
$(\Delta X_{\varphi})_{\mathrm{vac}} = 1$ 
for some angle $\varphi$. Since
\begin{equation}
X_{\varphi}^2 = 1 + 2 a^{\dagger} a + a^2 e^{-2 i \varphi} 
+ a^{\dagger 2} e^{2 i \varphi} ,
\end{equation}
the uncertainty $\Delta X_{\varphi}$ at any time $t$ can be 
determined by calculating the expectation values of $a$ and $a^2$ 
(recall that the mean number of photons 
$\langle a^{\dagger} a \rangle = |\alpha|^2$ is constant).
For $\beta = 0$, a straightforward calculation gives:
\begin{eqnarray}
\langle a \rangle = \langle a^{\dagger} \rangle^{\ast}
& = & |\alpha| \exp[ i (\zeta + \varphi_\alpha + \mu k^2)] 
\nonumber \\ & & \times
\exp\left[ - \dot{\mu} k^2 - |\alpha|^2 (1 - e^{2 i \mu k^2})
\right] ,
\end{eqnarray}
\begin{eqnarray}
\langle a^2 \rangle = \langle a^{\dagger 2} \rangle^{\ast}
& = & |\alpha|^2 \exp[2 i (\zeta + \varphi_\alpha + 2 \mu k^2)]
\nonumber \\ & & \times
\exp\left[ - 4 \dot{\mu} k^2 - |\alpha|^2 (1 - e^{4 i \mu k^2})
\right] .
\end{eqnarray}
Using these results, we can study numerically the behavior of the 
quadrature uncertainty. Also, in the relevant limits 
$|\alpha|^2 \gg 1$ and $\tau \ll 1$, an approximate expression is 
obtained
\begin{equation}
\label{eq:varX-appr}
(\Delta X_{\varphi})^2 \approx 1 + 2 |\alpha|^2 \left( 1 +
e^{-2 \Gamma} \cos 2\vartheta - 
2 e^{-\Gamma} \cos^2 \vartheta \right) ,
\end{equation}
\begin{eqnarray}
& & \Gamma = k^2 \tau^2 (1 + k^2 |\alpha|^2 \tau^4 /9) , \\
& & \vartheta = \zeta + \varphi_\alpha + k^2 |\alpha|^2 \tau^3 /3 
- \varphi .
\end{eqnarray}
The variance in Eq.~(\ref{eq:varX-appr}) is minimized for 
$\vartheta = n \pi$ ($n=0,\pm 1, \pm 2, \ldots$).
Then we find
\begin{equation}
  \label{eq:varX-appr-min}
(\Delta X_{\varphi})^2 \approx 1 + 2 |\alpha|^2 \left( 1 -
e^{-\Gamma} \right)^2 ,
\end{equation}
which is always larger than $1$.

Figures~4 and 5 show the basic features of the behavior of
the quadrature uncertainty. For simplicity, we took
$\zeta = \varphi_{\alpha} = 0$ in the numerical calculations.
In Fig.~4 the dependence of $\Delta X_{\varphi}$
on the homodyne phase $\varphi$ is shown for various field
intensities (the range of $\varphi$ from zero to $\pi$ is chosen 
because $\Delta X_{\varphi}$ is a periodic function of $\varphi$ 
with the period of $\pi$). For the parameters taken, the 
approximation (\ref{eq:varX-appr}) is excellent, and the 
quadrature uncertainty $\Delta X_{\varphi}$ reaches a minimum
for $\varphi$ about $k^2 |\alpha|^2 \tau^3 /3$ (i.e., for 
$\vartheta = 0$). The value of $\Delta X_{\varphi}$ at this
minimum is plotted in Fig.~5 versus $|\alpha|$ for several 
interaction times. Some discrepancy between exact and 
approximate results appears for small values of $\tau$, but
the approximation (\ref{eq:varX-appr-min}) is excellent for
field intensities large enough to assure $\Gamma > 1$. For
the parameters we considered, no squeezing of the field
quadrature was found.

\section{Discussion}

It should be clearly understood that we have no intention to
claim that the idealized system discussed in this paper might
be immediately used as a practical sensor of weak classical
forces (e.g., for detection of gravitational waves).
Our study is aimed at understanding the most basic properties of 
the quantum dynamics of the radiation field in a cavity with a 
movable mirror. The evolution of the field phase distribution,
studied in the framework of a fully quantum approach,
is of interest because it helps to clarify how the information
about an external force is carried by the light field in
much more complicated realistic measurement schemes. In fact,
it is very instructive to see how the evolution of the field
phase uncertainty determines the sensitivity of the scheme
and naturally implies the SQL on the measurement of the mirror 
position.

It should be emphasized that in our model we assumed that the
information about the phase shift accumulated by the cavity
field is read by means of a ``single-shot'' phase measurement.
Of course, this only gives information about the integrated
effect of an external force during the measurement time.
However, if one wants to determine the time dependence of the 
force, this will require a continuous monitoring of the field
phase. In a consistent quantum description, one should take
into account that the evolution of the system under the 
continuous measurement will be seriously affected by the 
measurement-induced state reduction. A model of continuous
broadband measurement for monitoring the position of a free
mass was recently analyzed in \cite{Mab98}. It would be also
interesting to have such a kind of quantum analysis for the 
continuous measurement on the system of the radiation field
interacting with a movable mirror.

\acknowledgements

CB thanks Kip S. Thorne, John Preskill, Yuri Levin, Bill Kells
and other participants of the Caltech QND reading group for useful 
and informative discussions. 
Financial support from the Lester Deutsch Fund is gratefully 
acknowledged. 
This work was supported in part by the Institute of Theoretical 
Physics at the Department of Physics at the Technion, and CB is 
grateful to the Institute for hospitality during his visit to 
the Technion. 
The LIGO Project is supported by the National Science Foundation 
under the cooperative agreement PHY-9210038.
AM was supported by the Fund for Promotion of Research at the 
Technion and by the Technion VPR Fund.

\end{multicols}

\twocolumn

%%%%%%%%%%%%%%%%%%%%%%%%%%%%%%%%%%%%%%%%%%%%%%%%%%%%%%%%%%%
\begin{figure}[htbp]
\label{fig:ph1}
\epsfxsize=0.45\textwidth
\vspace*{-31mm}
\centerline{\epsffile{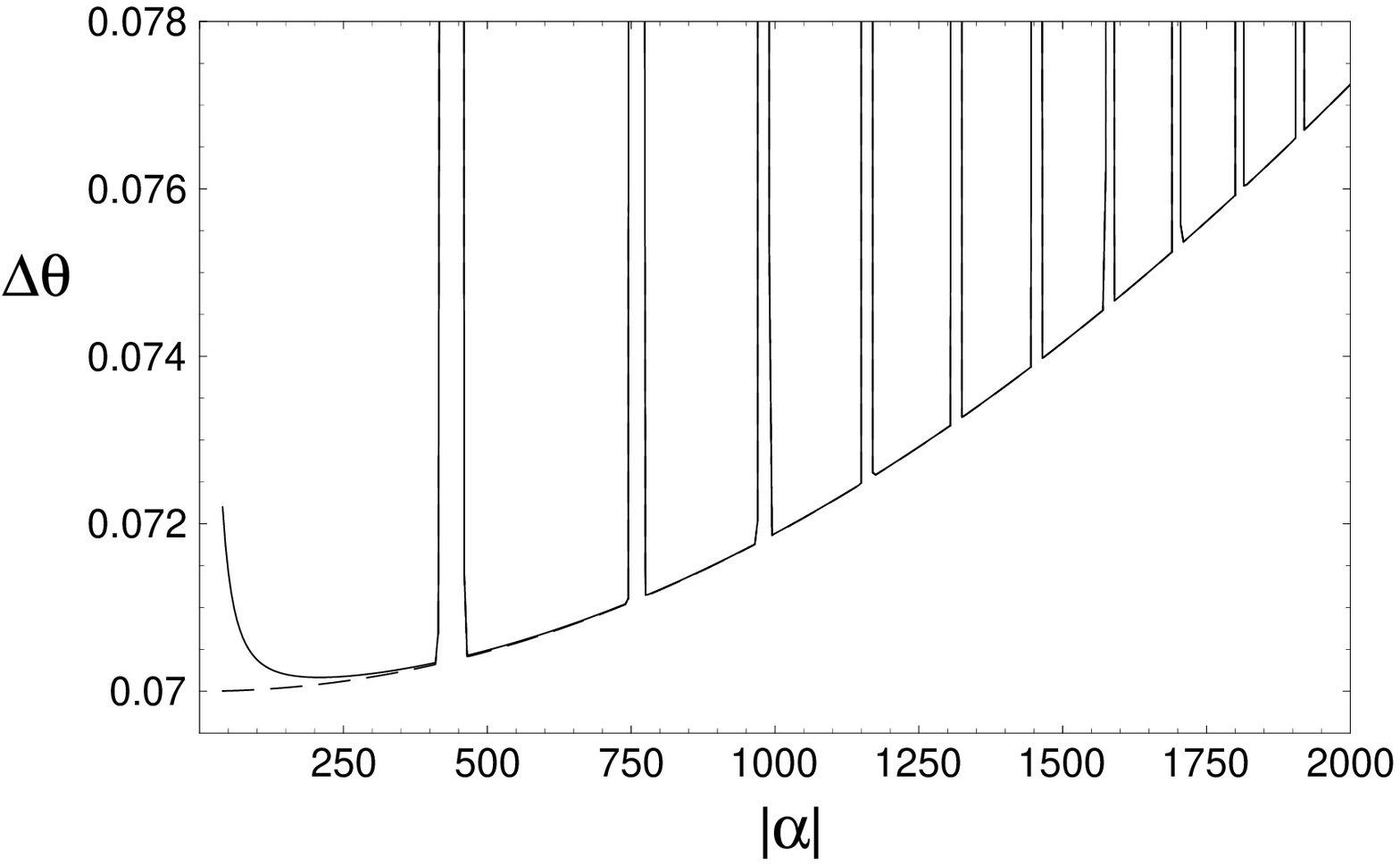}}
\vspace*{-25mm}
\caption{The phase uncertainty $\Delta \theta$
versus $|\alpha|$ for $k=7.0$ and $\tau = 0.01$. The exact result
is shown with the solid line and the approximate result, corresponding
to the distribution (\protect\ref{eq:PQt-appr}), is the dashed line. 
The difference between exact and approximate results 
appears only for $|\alpha| < 300$. If one ignores peaks, 
the corresponding curve coincides perfectly with the Gaussian 
dispersion $\sigma$.}
\end{figure}
%%%%%%%%%%%%%%%%%%%%%%%%%%%%%%%%%%%%%%%%%%%%%%%%%%%%%%%%%%%
%%%%%%%%%%%%%%%%%%%%%%%%%%%%%%%%%%%%%%%%%%%%%%%%%%%%%%%%%%%
\begin{figure}[htbp]
\label{fig:ph2}
\epsfxsize=0.45\textwidth
\vspace*{-30mm}
\centerline{\epsffile{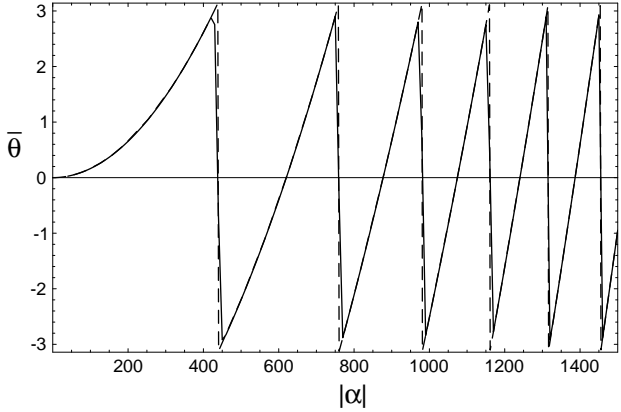}}
\vspace*{-25mm}
\caption{The mean phase $\bar{\theta}$ versus $|\alpha|$ for 
$k=7.0$ and $\tau = 0.01$. The exact result is shown with a solid 
line and the Gaussian approximation 
$\tilde{\theta} = \varepsilon_2$ with a dashed line. 
The difference between exact and approximate results is very 
insignificant. The mean phase $\bar{\theta}$ obtained from the
approximate distribution (\protect\ref{eq:PQt-appr}) is not 
shown here because it is almost indistinguishable from the exact 
value.}
\end{figure}
%%%%%%%%%%%%%%%%%%%%%%%%%%%%%%%%%%%%%%%%%%%%%%%%%%%%%%%%%%%
%%%%%%%%%%%%%%%%%%%%%%%%%%%%%%%%%%%%%%%%%%%%%%%%%%%%%%%%%%%
\begin{figure}[htbp]
\label{fig:ph3}
\epsfxsize=0.45\textwidth
\vspace*{-33mm}
\centerline{\epsffile{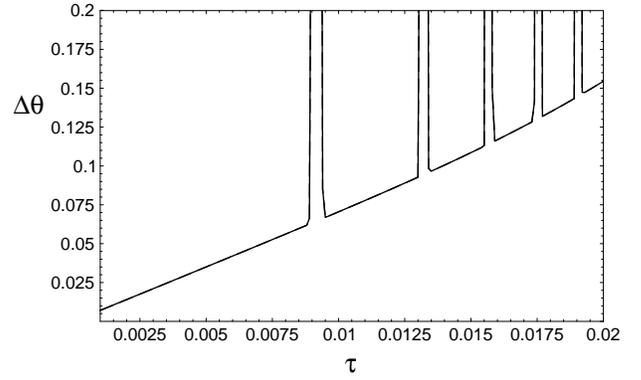}}
\vspace*{-25mm}
\caption{The phase uncertainty $\Delta \theta$ versus the scaled 
time $\tau$ for $k=7.0$ and $|\alpha| = 500$. The exact uncertainty 
and the approximate one, corresponding to the distribution 
(\protect\ref{eq:PQt-appr}), are indistinguishable. If one 
ignores peaks, the corresponding curve coincides perfectly with the 
Gaussian dispersion $\sigma$.}
\end{figure}
%%%%%%%%%%%%%%%%%%%%%%%%%%%%%%%%%%%%%%%%%%%%%%%%%%%%%%%%%%%
%%%%%%%%%%%%%%%%%%%%%%%%%%%%%%%%%%%%%%%%%%%%%%%%%%%%%%%%%%%
\begin{figure}[htbp]
\label{fig:sq1}
\epsfxsize=0.42\textwidth
\centerline{\epsffile{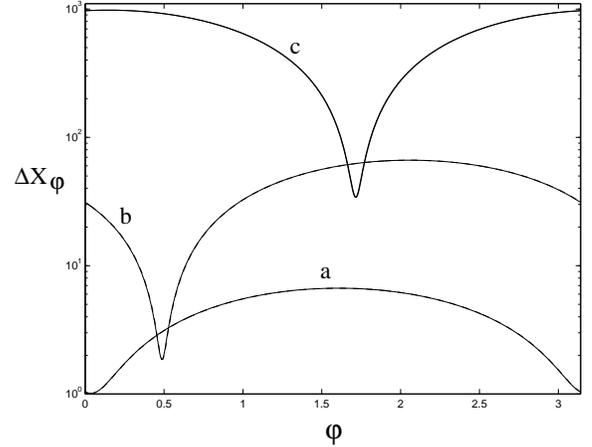}}
\vspace*{1mm}
\caption{The quadrature uncertainty $\Delta X_{\varphi}$
versus $\varphi$ for $k=3.3$, $\tau = 0.01$, and various $|\alpha|$:
(a) $|\alpha| = 10^2$, (b) $|\alpha| = 10^3$, (c) $|\alpha| = 10^4$. 
The difference between exact and approximate results (both of 
them shown in the plot) is almost invisible.}
\end{figure}
%%%%%%%%%%%%%%%%%%%%%%%%%%%%%%%%%%%%%%%%%%%%%%%%%%%%%%%%%%%
%%%%%%%%%%%%%%%%%%%%%%%%%%%%%%%%%%%%%%%%%%%%%%%%%%%%%%%%%%%
\begin{figure}[htbp]
\label{fig:sq2}
\epsfxsize=0.42\textwidth
\centerline{\epsffile{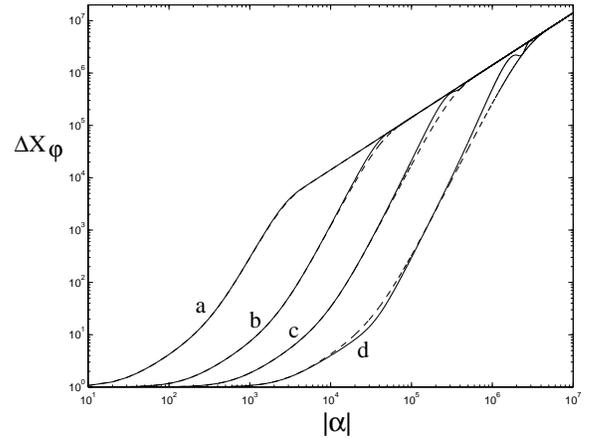}}
\vspace*{1mm}
\caption{The quadrature uncertainty $\Delta X_{\varphi}$ versus 
$|\alpha|$ for $k=3.3$, $\vartheta = 0$, and various $\tau$: 
(a) $\tau = 0.05$, (b) $\tau = 0.02$, (c) $\tau = 0.01$, 
(d) $\tau = 0.005$. Exact results are plotted with solid lines 
and approximate ones with dashed lines.}
\end{figure}
%%%%%%%%%%%%%%%%%%%%%%%%%%%%%%%%%%%%%%%%%%%%%%%%%%%%%%%%%%%

\end{document}